\magnification=1220
\baselineskip=15pt
\def\medskip{\vskip .2in}

\def\sma{\hskip 2.5mm}
\def\ds{\displaystyle}
\pageno=1
\footline{\hss\folio\hss}
\vskip2.5pc
\centerline{\bf Quantization of a Constant of Motion for the}
\centerline{\bf Harmonic Oscillator with a Time-Explicitly Depending Force } 
\vskip4pc
\centerline{G. L\'opez}
\vskip1pc
\centerline{Departamento de F\'isica de la Universidad de Guadalajara}
\centerline{Apartado Postal 4-137}
\centerline{44410~Guadalajara, Jalisco, M\'exico}
\vskip5pc
\centerline{December 1998}
\vskip1pc
\centerline{PACS~~~03.65.-w~~03.65.Ca~~03.65.Sq}

\vskip4pc
\centerline{ABSTRACT}
\vskip2pc
The quantization of a constant of motion for the harmonic oscillator with a
time-explicitly depending external force is carried out. This quantization
approach is compared with the normal  Hamiltonian quantization  approach.
Numerical results show that there are qualitative and quantitative differences
for both approaches, suggesting that the quantization of this constant of
motion may be verified experimentally.
\vfil\eject
\line{\bf I. Introduction\hfil}
\vskip1pc
Modern Physics has emerged as theories formulated in terms of Hamiltonians or
Lagrangians structures. Among them, one could mention classical physics, field
theories [1], quantum mechanics [2] and statistical mechanics [3]. Nature has
been kind enough to allows its description , so far, in terms of Hamiltonian
or Lagrangian  formulation, despite of some mathematical problems that these
formulations have by themselves [4]. These problems appear mainly  for
dissipative [5] or time-explicitly depending phenomena [6], where one  wonder
whether the associated constant of motion or Hamiltonian is the relevant
parameter to study the associated physics. For the harmonic oscillator with a
time-explicitly depending force (HOTED), it is possible to know a constant of
motion which is also time-explicitly depending [7]. Therefore, one may wonder
about any difference that HOTED could have from the quantum mechanics point of
view. In this paper, the quantization of a constant of motion fort HOTED is
associated and compared with the usual Hamiltonian approach.
\vskip3pc
\line{\bf II. Constant of Motion and Quantization Approach\hfil}
\vskip1pc
Consider that the harmonic oscillator is perturbed by the time-explicitly
depending force given by
$$f(t)=A\sin\Omega t\ ,\eqno(1)$$
where $A$ and $\Omega$ are the amplitude and the frequency of the force. The
resulting classical dynamical system can be written as
$${\ds dx\over\ds dt}=v\eqno(2a)$$
and
$${\ds dv\over\ds dt}=-\omega^2x+{A\over m}\sin\Omega t\ ,\eqno(2b)$$
where $\omega$ is the frequency of the free harmonic oscillations. A constant
of motion associated to [2] for $\Omega\not=\omega$ is given by [7] the function
$$K(x,v,t)={\ds m\over\ds 2}~(v^2+\omega^2x^2)+{\ds A\over\ds\Omega^2-\omega^2}~
\biggl(\Omega v\cos\Omega t+\omega^2 x\sin\Omega t~\biggr)
-{\ds A^2\over\ds 2m(\Omega^2-\omega^2)}~\sin^2(\Omega t)\ .\eqno(3)$$
One can readily see that (3) is a constant of motion since it satisfies the
following partial differential equation
$$v~{\ds\partial K\over\ds\partial x}+[-\omega^2x+{A\over m}~\sin\Omega t~]~
{\ds\partial K\over\ds\partial v}+{\ds\partial K\over\ds\partial t}=0\
.\eqno(4)$$
The approach used here to quantize (3) is based on the construction of the
associated Shr\"odinger's equation
$$i\hbar{\ds\partial\Psi\over\ds\partial t}=\widehat K(\hat x, \hat v,t)\Psi\
,\eqno(5)$$
where $\Psi$ is the wave function, $\Psi=\Psi(x,t)$, and $\widehat K$ is the
hermitian operator associated to function (3). This hermitian operator is
constructed by making the following substitutions
$$x\longrightarrow\hat x=x\sma\sma\hbox{and}\sma 
v\longrightarrow\hat v=-i{\hbar\over m}~{\ds\partial\over\ds\partial x}
\eqno(6)$$
which bring about the Shr\"odinger's equation
$$\eqalign{
i\hbar{\ds\partial\Psi\over\ds\partial t}&=
-{\ds\hbar^2\over\ds 2m}~{\ds\partial^2\Psi\over\ds\partial x^2}+
{1\over 2}m\omega^2x^2~\Psi\cr
&+\left\{-{\ds A^2\over\ds 2m(\Omega^2-\omega^2)}~\sin^2\Omega t+
{\ds A\omega^2 x\over\ds \Omega^2-\omega^2}~\sin\Omega t~\right\}\Psi\cr
&-i{\ds A\Omega\hbar\cos\Omega t\over\ds m(\Omega^2-\omega^2)}~
{\ds\partial\Psi\over\ds\partial x}\  .\cr}\eqno(7)$$
The first line on Eq. (7) represents the pure harmonic quantum oscillator.
The eigenfunctions of the pure harmonic oscillator are well known [8] and
are better handle with annihilation and creation operators, $\tilde a$ and 
$\tilde a^+$.  $\hat x$ and $\hat v$ are written in terms of these operators as
$$\hat x=\sqrt{\hbar\over\ds 2m\omega}~(\tilde a+\tilde a^+)\ ,\eqno(8a)$$
and
$$\hat v=-i\sqrt{\hbar\omega\over\ds 2m}~(\tilde a-\tilde a^+)\ .\eqno(8b)$$
Then, Eq. (3) can be written as
$$\eqalign{
\widehat K&=\widehat K_o(\tilde a^+,\tilde a)-{\ds A^2\over\ds
2m(\Omega^2-\omega^2)}\sin^2\Omega t\cr
&+{\ds A\over\ds \Omega^2-\omega^2}\sqrt{\hbar\omega\over\ds 2m}~\biggl\{
[\omega\sin\Omega t-i\Omega\cos\Omega t]\tilde a+[\omega\sin\Omega t+
i\Omega\cos\Omega t]\tilde a^+\biggr\}\ ,\cr}\eqno(9a)$$
where $\widehat K_o(\tilde a^+, \tilde a)$ is given by
$$\widehat K_o=\hbar\omega(\tilde a^+\tilde a+1/2)\ .\eqno(9b)$$
If $|n>$ represents an eigenfunction of the pure harmonic oscillator, $\hat
K_o$, one has the usual properties
$$\tilde a|n>\sqrt{n}~|n-1>\ ,\eqno(10a)$$
$$\tilde a^+|n>=\sqrt{n+1}~|n+1>\ ,\eqno(10b)$$
$$\tilde a^+\tilde a|n>=n|n>\eqno(10c)$$
and
$$[\tilde a, \tilde a^+]=1\ ,\eqno(10c)$$
where symbol $[ , ]$ represents the commutator between operators, $[\tilde a,
\tilde a^+]=\tilde a\tilde a^+-\tilde a^+\tilde a$. Now, proposing in (5) a
solution of the form
$$\Psi(x,t)=\sum_{n=0}^{\infty}c_n(t)|n>\ ,\eqno(11)$$
one gets the system of equations
$$\eqalign{
i\hbar\dot c_m&=\left[\hbar\omega(n+1/2)-{\ds A^2\over\ds 2m(\Omega^2-\omega^2)}
\sin^2\Omega t\right]~c_m\cr
&+{\ds A\over\ds\Omega^2-\omega^2}\sqrt{\hbar\omega\over\ds 2m}~
\biggl\{\sqrt{m+1}(\omega\sin\Omega t-i\Omega\cos\Omega t)c_{m+1}+
\sqrt{m}(\omega\sin\Omega t+i\Omega\cos\Omega t)~c_{m-1}\biggr\}\cr}\eqno(12)$$
which can be written in terms of the parameters
$$\tau=\omega t\ ,\sma\sma\epsilon=A/\hbar\Omega\
,\sma\sma\bar\hbar=\hbar/m\omega\sma\sma\hbox{and}\sma\sma
\rho=\Omega/\omega\eqno(13)$$
as
$$\eqalign{
ic_m'&=\left(n+1/2-{\ds\epsilon^2\over\ds 2\bar\hbar(1-\rho^2)}
\sin^2\rho\tau\right)~c_m\cr
&+{\ds\epsilon\ds\sqrt{2\bar\hbar}~(1-\rho)}
\biggl\{\sqrt{m+1}(\sin\rho\tau-i\rho\cos\rho\tau)c_{m+1}+
\sqrt{m}(\sin\rho\tau+i\rho\cos\rho\tau)c_{m-1}\biggr\}\ ,\cr}\eqno(14)$$
where $c_m'$ denotes the expression $c_m'=dc_m/d\tau$.

On the other hand, the Hamiltonian quantization approach for the system (2) is
formulated by the Shr\"odinger's equation
$$i\hbar{\ds\partial\Psi\over\ds\partial t}=\widehat H\Psi\ ,\eqno(15)$$
where the Hamiltonian operator $\widehat H$ is given by
$$\widehat H={\ds \hat p^2\over\ds 2m}+{1\over 2}m\omega^2x^2-xA
\sin\Omega t\ .\eqno(16)$$
The operators $\hat x$ and $\hat p$ can be written in terms of the annihilation 
and creation operators, $a$ and $a^+$, as
$$\hat x=\sqrt{\hbar\over\ds 2m\omega}~(a+a^+)\eqno(17a)$$
and
$$\hat p=-i\sqrt{\hbar\omega m\over\ds 2}~(a-a^+)\ ,\eqno(17b)$$
where $a$ and $a^+$ satisfy the same relations (10). Written (16) in terms of 
these operators, it follows that
$$\widehat H=\widehat H_o(a^+, a)-\sqrt{\hbar\over\ds 2m\omega}~ A\sin\Omega t~(a+a^+)\
,\eqno(18)$$
where $\widehat H_o(a^+, a)$ is given by $\widehat H_o=\hbar\omega(a^+a+1/2)$. Using
the same expansion (10) and parameters (13), one gets the system of equations
$$ic_m'=(m+1/2)c_m-\sqrt{\bar\hbar\over\ds 2}~\epsilon\rho\sin\rho\tau~
(\sqrt{m}~c_{m-1}+\sqrt{m+1}~c_{m+1})\ .\eqno(19)$$
\vskip3pc
\line{\bf III. Numerical Solution\hfil}
\vskip1pc
To know the evolution of $c_m(\tau)$ for Eqs. (14) and (15), these can be solve
numerically. The fixed parameters used are $\bar\hbar=0.4$ and $\rho=6.25$, and
one studies the dependence of the system on the parameters $\epsilon$ and
$\tau$. The initial conditions are 
$$c_o(0)=1,\sma\sma,c_j(0)=0\sma\hbox{for}\sma j=1,2,\dots\eqno(20)$$

The Fig. 1 shows the evolution of the probabilities $|c_o(\tau)|^2$ and
$|c_1(\tau)|^2$ and their dependence on the parameter $\epsilon$. The upper
plots correspond to the quantization of the constant of motion (Eq. 14), and the
lower plots correspond to the quantization of the Hamiltonian (Eq. 19). For the
same $\epsilon$, the peak values of the probabilities $|c_o(\tau)|^2$ and
$|c_1(\tau)|^2$ occur at different times ($\tau$) and with different
amplitudes. 

The Fig. 2 shows the maximum number of exited states involved in the dynamics
of the system for a given $\epsilon$. The criterion used is the following: one
defines that a state $|n>$ is involved in the dynamics if the probability that
the system to be in this state is higher than $0.0001$, $|c_n(\tau)|^2>0.0001$.
As one can see, for a given $\epsilon$ the number of levels involved in the
Hamiltonian quantization can be much larger than the levels involved in the
constant of motion quantization.

The Fig. 3 shows the evolution of the expected value of $x^2$,
$$<x^2>=<\Psi|x^2|\Psi>=\sum_{m,n=0}^{\infty}c_m^*(\tau)c_n(\tau)<m|x^2|n>\
,\eqno(21)$$
for the constant of motion and Hamiltonian quantization. As one can see, the
value for the Hamiltonian case is about one order of magnitude higher than its
value for the constant of motion case.
\vskip3pc
\line{\bf III. Conclusions\hfil}
\vskip1pc
The quantization of the harmonic oscillator with a time-explicitly depending
external force has been studied from the point of view of the constant of
motion. The results have been compared with the known case of Hamiltonian
approach. The probabilities $|c_o(\tau)|^2$ and $|c_1(\tau)|^2$, the
expected value $<x^2>$ and the number of exited states involve in the
dynamics are quite different for the constant of motion and Hamiltonian 
approaches. These results may be checked experimentally taking care that the
external force added to the harmonic oscillator system is of the form given by
Eq. (1).
\vfil\eject
\vskip3pc
\line{\bf References\hfil}
\vskip1pc
\obeylines{
1. L.D. Landau and E.M. Lifshitz,``The Classical Field Theory,"
\quad  Addison-Wesley Publishing Co., Inc. 3rd ed., 1970.
2. P. A. M. Dirac, ``The Principles of Quantum Mechanics," 3rd ed., 
\quad Claredon Oxford, 1947.
3. K. Huang, ``Statistic Mechanics," John-Wiley, New York, London, 1963.
4. J. Douglas, Trans. Amer. Math. Soc., {\bf 50} (1941)71.
5. G. L\'opez, Ann. of Phys., {\bf 251}, No.2(1996)372.
6. G. L\'opez and J.I. Hern\'andez, Ann. of Phys. {\bf 193}(1989)1.
7. G. L\'opez, Int. Jour. of Theo. Phys. {\bf 37}, No.5(1998)1617.
8. G. L\'opez, IL Nuovo Cimento B, May,(2000). math-ph/9910008.
9. A. Messiah, ``Quantum Mechanics," John Wiley and Sons, Inc., Vol.I, 1958.
}
\vfil\eject
\leftline{\bf Figure Captions}
\vskip1pc

Fig.1 Evolution of the probabilities $|c_o(\tau)|^2$ and $|c_1(\tau)|^2$ for
the system (14) and (19). The value of the other parameters are
$\bar\hbar=0.4$ and $\rho=6.25$. The initial conditions are given by Eq. (20).
\vskip1pc
Fig. 2 Number of exited states involved in the dynamics as a function of the
parameter $\epsilon$. The other parameters are the same as Fig. 1.
\vskip1pc
Fig. 3 Evolution of the expected value $<x^2>$ for the constant of motion and
Hamiltonian quantization approaches. The parameter epsilon takes the values
$\epsilon=0,5,10$. The other parameters and initial conditions are the same as
Fig. 1.

\end